# Probing ultrafast carrier dynamics and nonlinear absorption and refraction in core-shell silicon nanowires

Sunil Kumar,[1,a)] M. Khorasaninejad,[2] M. M. Adachi,[2] K. S. Karim,[2] S. S. Saini,[2] and A. K. Sood[1]

[1]*Department of Physics and Center for Ultrafast Laser Applications, Indian Institute of Science, Bangalore 560012, India*
[2]*Department of Electrical and Computer Engineering, University of Waterloo, Canada N2L 3G1*
[a)]*Corresponding author. Email: skumar@physics.iisc.ernet.in*

**Abstract:** We investigate the relaxation dynamics of photogenerated carriers in silicon nanowires consisting of a crystalline core and a surrounding amorphous shell, using femtosecond time-resolved differential reflectivity and transmission spectroscopy at photon energies of 3.15 eV and 1.57 eV. The complex behavior of the differential transmission and reflectivity transients is the mixed contributions from the crystalline core and the amorphous silicon on the nanowire surface and the substrate where competing effects of state filling and photoinduced absorption govern the carrier dynamics. Faster relaxation rates are observed on increasing the photo-generated carrier density. Independent experimental results on crystalline silicon-on-sapphire help us in separating the contributions from the carrier dynamics in crystalline core and the amorphous regions in the nanowire samples. Further, single beam z-scan nonlinear transmission experiments at 1.57 eV in both open and close aperture configurations yield two-photon absorption coefficient $\beta$ (~3 cm/GW) and nonlinear refraction coefficient $\gamma$ (-2.5x$10^{-4}$ $cm^2$/GW).

**Keywords:** Silicon, Nanowires, Femtosecond, Ultrafast spectroscopy, Nonlinear optical coefficients

**PACS:** 78.40.Fy ; 78.67.Uh ; 78.47.J-

## I. INTRODUCTION

Energy relaxation processes play an important role in hot carrier transport in semiconductor based fast electronic and optoelectronic devices. Spatial confinement of charge carriers in semiconducting nanostructures not only leads to band gap enhancement [1] but also gives rise to surface induced charge states in the forbidden gap. Silicon nanowires are being used to fabricate multi-devices such as transistors, photo-emitters and photo-sensors, all on a single chip [2-4], mainly due to the ease of large scale fabrication and integration [5]. Crystalline core-amorphous shell silicon heterostructures have been recently used in applications including battery electrodes [6], electrical switching [7] and solar cells [8]. The advantage in using core-shell nanowires over purely crystalline or amorphous silicon nanowires is the passivation of crystalline silicon to be used as an efficient electrical conducting pathway [6] together with high optical absorption by amorphous silicon, making them suitable for solar cell applications [9]. Ultrafast time-resolved pump-probe spectroscopy has contributed in understanding the energy relaxation processes in various semiconducting nanowires [10-13]. Recently, pump-probe spectroscopy was reported on randomly networked crystalline silicon nanowires (SiNWs) on a sapphire substrate where the electron relaxation rate was found to be dependent on the nanowire diameter and the energy of the probed states within the forbidden energy gap [14]. Inspired by these studies, interest was generated to carry out the present study on crystalline core-amorphous shell silicon nanowires for the carrier relaxation dynamics and nonlinear optical properties for its potential applications in silicon based visible light modulators and all-optical switching devices [15].

We performed femtosecond optical measurements using 70 fs laser pulses, on preferentially vertically grown crystalline core-amorphous shell silicon nanowires on a glass substrate. Electron-hole pairs are optically injected into the nanowires and the subsequent temporal evolution of their population distributions are studied with femtosecond time resolution in the pump-probe reflection and transmission experiments. We examine the dependence of the photogenerated carrier dynamics on the pump-photon energy (3.15 eV and 1.57 eV) and the pump-fluence. The transient differential reflectivity ($\Delta R/R$) and transmission ($\Delta T/T$) signals have contributions from both the crystalline core and the amorphous silicon shell on the nanowire surface, leading to a complex behavior of the transients that reveal multiple relaxation components of the carrier dynamics with time constants ranging from sub-ps to many hundreds of ps governed by the photoinduced absorption and the state filling effects. Independent experiments performed under same experimental conditions on crystalline silicon-on-sapphire (SOS), on the other hand, reveal only single exponential decay of the photoinduced transmission transients with time constant of ~ 500 ps. Also, we have estimated two photon absorption coefficient $\beta \sim 3$ cm/GW and negative nonlinear refractive index $\gamma \sim -2.5 \times 10^{-4}$ $cm^2$/GW from z-scan experiments on the nanowires at 790 nm.

## II. EXPERIMENTAL

The core-shell silicon nanowires were fabricated by vapor-liquid-solid method in a plasma-enhanced-chemical-vapor-deposition system on Corning 1737 glass substrate [9]. A 2 nm thick layer of Sn was deposited by e-beam evaporation followed by an anneal at 400 C to form Sn droplets used as the catalyst for nanowire growth. Silane was used as the source gas and nanowire growth temperature was 400 C. High resolution electron microscopy showed that the crystalline core has a diameter of about 10 nm which is surrounded by an amorphous silicon shell [9] resulting in a total nanowire average diameter of ~60 nm as seen from the scanning electron microscopic topograpy in Fig. 1(a). Raman spectrum of the nanowire sample shows a sharp peak at 520 cm$^{-1}$ related to the highly crystalline core and a broad peak at 480 cm$^{-1}$ due to scattering from amorphous silicon on the nanowires and on the substrate [9]. The overall thickness of the silicon nanowire film was ~600 nm measured independently by using atomic force microscopy providing an estimate of the nanowire length.

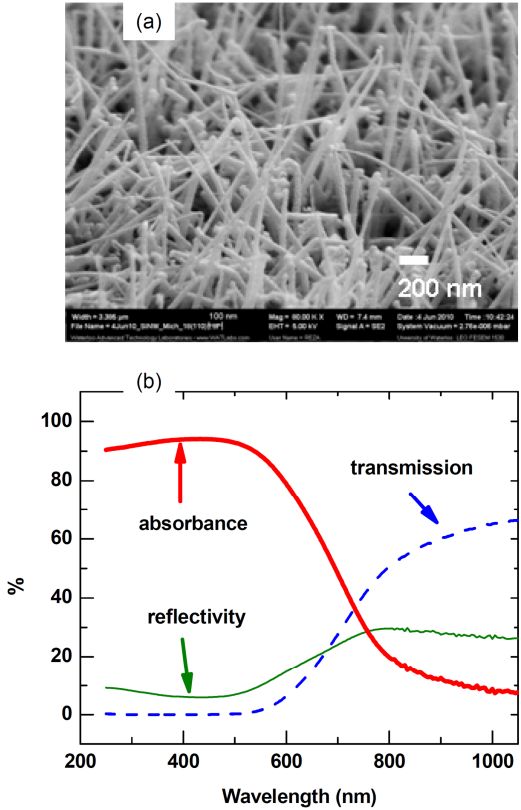

**FIG. 1.** Characterization of the silicon nanowires deposited on 1767 glass [15]. (a) Scanning electron microscopic topography for estimating the average total diameter of about 60 nm of the core-shell silicon nanowires, (b) linear optical absorption measurement using a visible-infrared absorption spectrometer.

The linear optical reflectivity and transmission data close to normal incidence obtained using the conventional visible-IR spectrophotometer, are shown in Fig. 1(b) along with the calculated absorbance. The band gap of amorphous silicon (a-Si) is around 1.78 eV (~700 nm) whereas the absorption from the nanowire sample is extended much beyond this wavelength and can originate mostly from the crystalline silicon core [9]. The linear absorbance at 395 nm is about 95% which corresponds to an absorption coefficient α ~ 4.5x10$^4$ cm$^{-1}$, ~14 times higher than at 790 nm.

Femtosecond laser pulses centered at 790 nm (1.57 eV) from a Ti:sapphire regenerative amplifier providing laser pulses at 1 kHz repetition rate and at 395 nm (3.15 eV) generated in a beta barium borate crystal, were used in the present degenerate and nondegenerate pump-probe experiments in a noncollinear geometry. The laser beam size on the sample was larger than 1 mm which results in averaging over a large number of nanowires. For pump-fluence dependent measurements we used neutral density filters in the pump beam path. We examined the dependence of pump-photon energy (1.57 eV and 3.15 eV) and pump-fluence on the carrier dynamics by measuring the time resolved photo-induced changes in sample transmission (ΔT) and reflection (ΔR) normalized with respect to their normal values (T and R) at 1.57 eV probe with probe-fluence of ~45 μJ/cm$^2$. All measurements were performed at room temperature with pump and probe polarizations crossed with each other and an analyzer crossed with the pump polarization was placed at the detector. For the nonlinear transmission single beam z-scan measurements at 790 nm, we used a focusing lens (focal length 15 cm) to continuously vary the laser power on the sample from 2 MW/cm$^2$ to ~30 GW/cm$^2$.

In our experiments the probe pulse centered at 1.57 eV is in the band gap region of a-Si where there are no optically active energy levels as suggested from the strong suppression of optical absorption beyond 1.78 eV [9]. A pump-probe signal at this probe energy thus signifies a finite absorption at 1.57 eV mostly due to phonon-assisted indirect transitions in the crystalline-silicon (c-Si) core and possibly a contribution from the band-tail states in a-Si. Moreover, the diameter of the crystalline core is ~10 nm where quantization effects are weak.

## III. RESULTS AND DISCUSSION

Differential transmission transients (ΔT/T) from the SiNWs at 3.15 eV pump collected at various pump-fluences are presented in Fig. 2(a). The recovery of the signals is composed of three components with negative and positive amplitudes and time constants from subps to ns arising from competing photoinduced absorption and state filling effects. A sum of exponentially decaying functions $\sum_k (1 - e^{-t/\tau_r}) A_k e^{-t/\tau_k}$ has been used to capture the complete dynamics of the photo-injected carriers in the SiNWs. Here the first term inside the brackets is used to capture the initial rise time ($\tau_r$) of the signal and corresponds to the thermalization of the photocarriers by carrier-carrier and/or carrier-optical phonon scattering, and $A_k$ and $\tau_k$ are the amplitudes and decay times of various components in the signal as displayed in Fig. 2b. The fits have been shown by thick continuous lines in Fig. 2a. We clearly see two negative components, one very fast with relaxation time $\tau_1 \sim 1.6$ ps and another very slow with $\tau_3 > 1$ ns attributed to the photoinduced absorption of carriers whereas at intermediated times we have a positive component with relaxation time constant $\tau_2 \sim 90$ ps that can arise from state filling effects induced saturation of carrier population in the energy states at the probe photon energy thus blocking the normal absorption of the probe photons in transition from the valence band to the conduction band. The long lived negative component could be a result of photo-induced absorption of the probe photons from a set of low energy defect states to the conduction band before the carrier recombination takes place to reach a complete equilibrium. Radiative recombination lifetime varying from picoseconds in defect-free and tens of nanoseconds through defect states in SiNWs have been suggested from time-resolved photoluminescence studies [16].

The pump-fluence dependence of each component is shown in Fig. 2b where the linear increase in the amplitudes and faster relaxation times (except $\tau_3$) can be seen at higher fluences. The rise time $\tau_r \sim 150$ fs is found to be fluence-independent. In nanowires, the carrier-carrier interactions are larger due to the spatial confinement of carriers as compared to bulk semiconductor, hence, lead to dominant Augur recombination process in the relaxation [14] according to which the relaxation times decrease with the increasing fluence. Also, at high fluences, the photogenerated carrier density can become sufficiently high to cause band-gap renormalization [14] and hence pronounced effects in the carrier dynamics at high fluences. Generally, Auger recombination mechanism is suggested to operate at carrier densities larger than $10^{21}$ cm$^{-3}$ [17]. However, the maximum pump-fluence at 395 nm that is used in our experiments is 620 $\mu$J/cm$^2$, corresponds to a photogenerated free carrier density N $\sim 5.4 \times 10^{19}$ cm$^{-3}$ where Auger recombination mechanism is not expected for the carrier relaxation. Though the decrease in the relaxation times $\tau_1$ and $\tau_2$ is observed at 395 nm pump, the same is found to be absent when 790 nm pump was used (discussed later). Moreover, in a-Si the carrier relaxation time as fast as 0.6 ps was reported to be fluence-independent even at carrier densities much higher than ours [17]. On the other hand, in crystalline semiconducting nanowires, fluence dependent relaxation times have been observed where Auger process is involved to play a role [10,14]. These observations may suggest that the origin of $\tau_1$ and $\tau_2$ is not simply related to either the carrier relaxation in a-Si or c-Si in the nanowire core separately; rather the presence of one affects the other.

We believe that in the present heterostructure SiNWs, the crystalline core and the amorphous silicon on the nanowire surface both contribute to the observed decay of the signals. The optical penetration depth ($\sim 1/\alpha$) at 395 nm is about 220 nm and much larger ($\sim 3$ micron) at 790 nm which effectively means that the photon absorption (at both 395 nm and 790 nm) takes place not only in the amorphous surface regions of the nanowires but also inside the crystalline core creating electron-hole pairs. Therefore, both the crystalline and amorphous silicon are contributing to the measured transients. We may emphasize that much of the incident light is transmitted to the substrate but the glass or the sapphire alone does not show any pump-probe signal [18]. Very slow ($\sim$ns) relaxation time in crystalline bulk silicon is commonly known [19,20] whereas much faster relaxation times from tens of ps to subps have been reported for a-Si [17,18,21].

To decipher the role of the crystalline core from that of the amorphous silicon, we carried out experiments under the same experimental conditions on commercially available crystalline silicon-on-sapphire (SOS) (0.4 micron crystalline silicon film on 0.5 mm sapphire) at 395 nm and 790 nm pump. The results are displayed in Fig. 3. In this case we observe a single exponential decay of the photocarriers with relaxation time constant $\tau \sim 500$ ps, independent of the excitation wavelength and the pump-fluence or the photogenerated carrier density. The only difference in the carrier dynamics at the two pump wavelengths is in their population build up immediately after the pump excitation (inset of Fig 3). At 395 nm, the carriers are injected into states in the conduction band far off from the Fermi energy and hence take more time ($\tau_r \sim 0.9$ ps) to reach the states connected by the probe (790 nm) than they take in the case of excitation with the 790 nm pump ($\tau_r \sim 0.4$ ps).

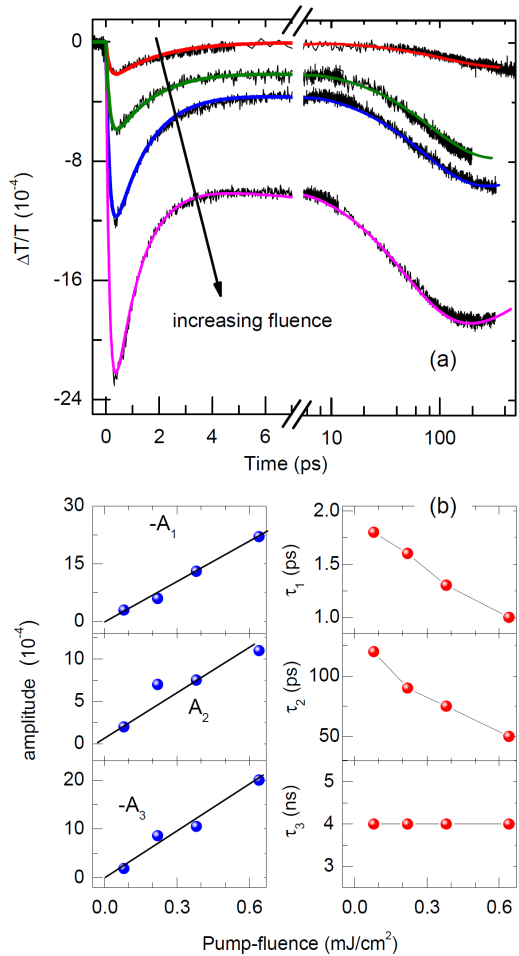

**FIG. 2.** Differential transmission transients from the SiNW sample measured at various pump-fluences with pump excitation at 3.15, (a) the raw data with logarithmic x-axis at longer times shown for clarity and (b) the fluence-dependence of amplitude (A) and time-constant ($\tau$) of three exponentially decaying relaxation components of the signals.

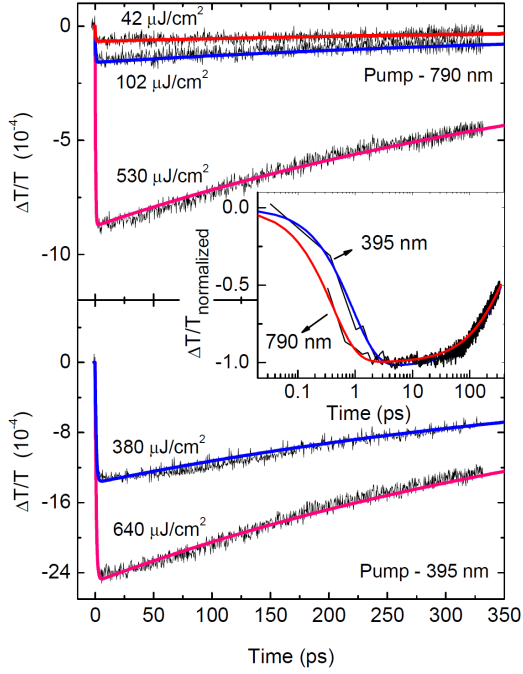

**FIG. 3.** Differential transmission transients from SOS measured using pump excitation at 790 nm (top panel) and 395 nm (bottom panel) at various pump-fluences. The Continuous lines are fits using a rising contribution and a single exponential decay. The inset displays the semi-logarithmic plot to clearly show the difference in the rising time of the signals at the two pump wavelengths.

From the above understanding, the initial fast negative component with decay time constant $\tau_1 \sim 1.6$ ps can be attributed to the photoinduced absorption where the pump-injected carriers in the excited states undergo secondary excitation by the probe photons and free carriers are contributed in the amorphous regions (band gap 1.78 eV). In crystalline silicon the lowest band gap (indirect) energy is 1.12 eV and the band gap for direct transition is much larger than the single photon energies used in our experiments. Amorphous silicon, on the other hand, with direct band gap energy of 1.78 eV [9] is highly absorbing at photon energies above 1.78 eV because momentum is not conserved during interband transitions due to atomic disorder [22]. Also, electrons and holes by dangling bond defects occur in amorphous silicon [23]. Therefore a pump-probe signal at probe photon energy of 1.57 eV which is in the mid gap region for both the crystalline and amorphous silicon effectively means that the sample has a finite absorption at 790 nm which get affected by the pump excitation. The probe can be absorbed through indirect (phonon assisted) transition in the crystalline silicon and electronic transitions to the band tail states in the amorphous silicon. The slow relaxation components ($\tau_2 \sim 100$ ps and $\tau_3 > 1$ns) are due to the relaxation of the photogenerated carriers in the crystalline core region of the nanowires. The slower one ($\tau_3$) is nearly fluence independent while $\tau_2$ is fluence dependent (Fig. 2b). The later can be related to the carrier trapping in the crystalline-amorphous interface of the core-shell nanowires.

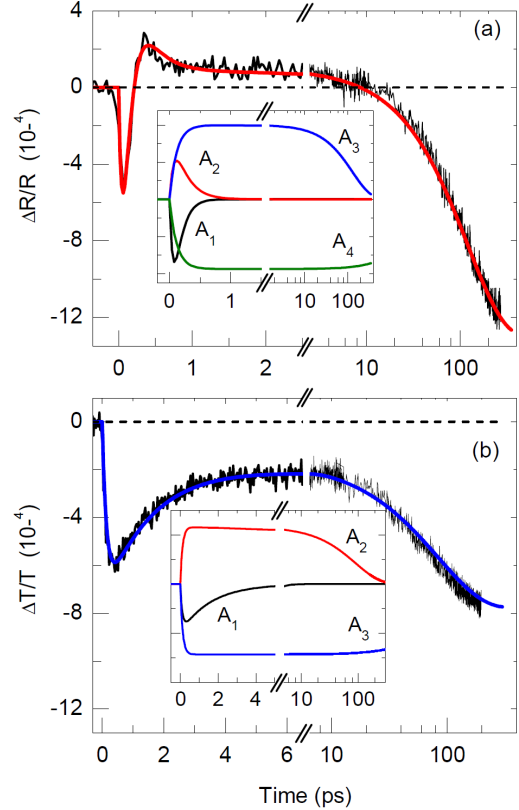

**FIG. 4.** Differential reflectivity and transmission transients from core-shell SiNWs using pump excitation at 3.15 eV with fits shown by thick continuous lines, using a sum of exponentially decaying functions. The insets show four (three) individual components used in fitting the $\Delta R/R$ ($\Delta T/T$) signal. Note that for clarity of presentation, the x-axis is linear at short times and logarithmic after the break.

Interestingly, when the pump-probe signal was collected in the reflection geometry ($\Delta R/R$), we observed a clear difference in the signal behavior at faster times (below 5 ps). Fig. 4 displays the results for this comparison between the $\Delta R/R$ and $\Delta T/T$ signals measured using at 395 nm pump using a pump-fluence of ~220 µJ/cm². The initial decrease in both the signals just after photo-excitation takes about 100 fs in $\Delta R/R$ and ~150 fs in $\Delta T/T$ to reach the minimum. At longer times larger than ~20 ps both the signals ($\Delta R/R$ and ΔT/T) behave same way however difference in their behavior can be seen at shorter times where single negative component in ΔT/T but one negative and one positive components in the ΔR/R signal are visible. It appears that the fast negative component in the ΔT/T has two contributions one from the amorphous silicon on the nanowire and another from that on the substrate. These get separated out in the ΔR/R geometry due the change in the polarity of the signal-component contributed by the residual amorphous silicon on the substrate.

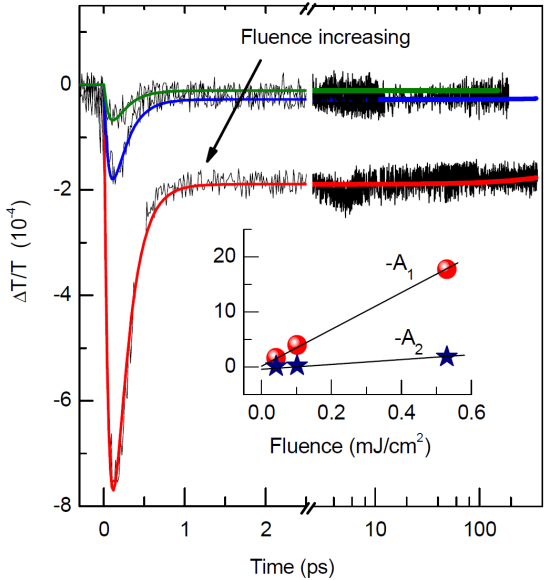

**FIG. 5.** Differential transmission transients for SiNWs using 790 nm pump at various pump-fluences. The inset shows the linear fluence dependence of the magnitudes of two individual relaxation components in the signal but the time-constants remain fluence independent.

The pump-probe signal immediately after the pump excitation is a consequence of cascaded absorption of a pump photon and a probe photon in nonresonant experiments like ours. In that case the magnitude of the signal should be smaller at 790 nm pump when compared with that at 395 nm pump because of very small density of states near the band tails of amorphous silicon band gap (700 nm). This is indeed the case as we have seen for the SOS (Fig. 3). To verify this point further, we pursued degenerate pump-probe measurements at 790 nm on the SiNWs as well. The results are displayed in Fig. 5. We again see the initial photoinduced absorption with a rise time of $\tau_r \sim 120$ fs, nearly identical to that at 395 nm pump. As it can be seen, the magnitude of the signal near the zero delay is much smaller as compared to the 395 nm pump excitation (Fig. 2) at the same pump-fluence. This is expected due to much smaller absorption in the crystalline core as well as the band tails of the amorphous silicon at 790 nm. Here, the combined absorption of one pump and one probe photon is responsible for the sharp dip at the zero delay. A positive component with decay time of ~10 ps could be identified only at the highest experimental pump-fluence (Fig. 5) and the data at all the fluences could be fitted using only two negative components with decay times of $\tau_1 \sim 200$ fs and $\tau_2 \sim 4$ ns. The relaxation times are found roughly fluence-independent unlike the 395 nm pump as discussed before and the linear dependence of the amplitudes $A_1$ and $A_2$ can be seen in the inset of Fig. 5.

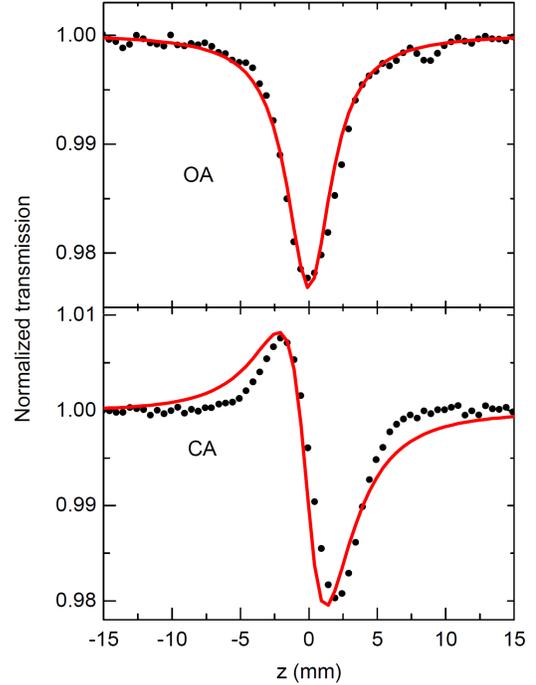

**FIG. 6.** Open aperture and close aperture z-scan results from silicon nanowire film at 790 nm. The continuous lines are theoretical fits to estimate the two photon absorption coefficient and nonlinear refractive index.

The photoinduced absorption near the zero delay in the degenerate pump-probe signal at 790 nm (Fig. 5) can be used to estimate the two photon absorption coefficient β of SiNWs from the relation [24] $\beta = \ln(1 + \Delta T_{max}/T)/LI_{pump}$. At the pump fluence of 530 μJ/cm$^2$ β is ~ 1.3 cm/GW. This is compared with the results obtained from open aperture (OA) and close aperture (CA) z-scan measurements. Optical limiting behavior arising from two photon absorption at 790 nm in the OA and negative refractive nonlinearity (γ) in the CA z-scan results can be seen in Fig. 3(b). Consistent theoretical fits [25] to the OA and CA data using $\alpha(I) = \alpha_0 + \beta I$, $n(I) = n_0 + \gamma I$ where $\alpha_0$ and $n_o$ are the linear absorption and refraction coefficients, give β ~ 3 cm/GW and γ ~ -2.5 x $10^{-4}$ cm$^2$/GW. The value of γ is small but β is significant when compared with highly nonlinear materials such as graphene and BCN [24,25] in the visible range. Similar value of β (~ 4 cm/GW) and a smaller but negative nonlinear refractive index was reported previously for a-Si waveguides [21] which are much higher than those for the c-Si [26].

## IV. CONCLUSIONS

To summarize, we have studied the photoinduced carrier dynamics in core-shell silicon nanowires where we observe the combination of fast carrier relaxation within 1.6 ps at 395 nm pump and 0.2 ps at 790 nm pump in the amorphous surface, and very slow relaxation larger than 100 ps in the crystalline core region of the nanowires. The large values of optical nonlinearities estimated from z-scan measurements along with the ultrafast response of better than 1.6 ps mainly in the amorphous silicon shell make the core-shell SiNWs a good candidate for potential applications in silicon based visible light modulators and all-optical-switching devices.

**ACKNOWLEDGEMENTS:** AKS and SK acknowledge the financial support from Department of Science and Technology, India.